\documentclass[aps,graphicx,twocolumn]{revtex4}
\usepackage{amsmath}
\usepackage{amscd}
\usepackage{graphicx}
\begin{document}

\title{Faithful qubit transmission against collective noise without ancillary qubits
\footnote{Published in \emph{Applied Physics Letters} \textbf{91},
144101 (2007)}}
\author{ Xi-Han Li,
 Fu-Guo Deng,\footnote{Author to whom correspondence should be addressed. Also at: Department of Physics, Beijing Normal University, Beijing
100875,  China; Electronic mail: fgdeng@bnu.edu.cn} and Hong-Yu
Zhou}
\address{The Key Laboratory of Beam Technology and Material
Modification of Ministry of Education, and Institute of Low Energy
Nuclear Physics, Beijing Normal University, Beijing 100875,  China}
\date{\today }

\begin{abstract}
We present a faithful qubit transmission scheme with linear optics
against collective noise, not resorting to ancillary qubits. Its
set-up is composed of three unbalanced polarization interferometers,
based on a polarizing beam splitter, a beam splitter and a half-wave
plate, which makes this scheme more feasible than others with
present technology. The fidelity of successful transmission is 1,
independent of the parameters of the collective noise, and the
success probability for obtaining an uncorrupted state can be
improved to 100\% with some time delayers. Moreover, this scheme has
some good applications in one-way quantum communication for
rejecting the errors caused by the collective noise in quantum
channel.
\end{abstract}\maketitle

The main task of quantum communication is transmitting and
exchanging quantum information between two remote parties
\cite{rmp}. In the last two decades, quantum communication had a
drastic progress both in theory and experiment. In laboratory, the
polarization state of photons is usually chosen as the qubit for
quantum communication due to its maneuverability. For example, the
famous BB84 \cite{bb84} quantum key distribution (QKD) scheme uses
two conjugate polarization bases of photons to create a secret key,
and the first quantum teleportation \cite{93} scheme selects the
polarization states of photons to teleport an unknown quantum state.
However, the polarization freedom of photons is incident to be
influenced by the thermal fluctuation, vibration and the
imperfection of the fiber, i.e., the noise in quantum channel. This
is a serious obstacle to the application of quantum communication in
practice. Then, various error correction and error rejection methods
are proposed. For instance, Walton et al. \cite{subspace} proposed a
scheme for rejecting the errors introduced by noise with
decoherence-free subspaces \cite{space}. Quantum redundancy code
\cite{code} is introduced by entangling the signal with some
ancillary qubits before the transmission. After the transmission,
the ancillary qubits will be measured, and some unitary operations
will be performed on the signal to correct the errors according to
the measurement results. Theoretically, the number and the kind of
errors that can be corrected depend on the number of the ancillary
qubits and the kind of code. In a long-distance quantum
communication, entanglement purification \cite{purify} is introduced
to decrease the influence arisen from the noise, with which a higher
fidelity state is obtained by sacrificing several qubits. On the
other hand, the steps to get a genuine pure entangled state are
always infinity and the cost of resource grows with the distance
between the two parties.

In 2005, Kalamidas \cite{correction} proposed two linear-optical
single-photon schemes to reject and correct arbitrary qubit errors
without additional particles. The first one obtains an uncorrupted
qubit at a definite time of arrival. Its success probability is 0.5
when varying the noise parameters over their entire ranges;
otherwise, the probability depends on the parameters of noise. The
second scheme is based on self-correcting and its success
probability is 1. That is, the receiver can always get an
uncorrupted state. Later Han \emph{et al.} \cite{mix} proved that
these two schemes are also suitable for mixed state. However, in
these two protocols, at least two fast polarization modulators
(Pockels cell), whose synchronization makes it difficult to be
implemented with current technology
 \cite{coherent}, are employed. Subsequently, de Brito and Ramos \cite{coherent} presented a
different setup to realize error correction without Pockels cells,
in which coherent states were used instead of single-photon pulses.
They also use Faraday mirror and common mirror substituting the half
wave plate. This new setup is passive and suitable for bright
coherent states, but due to the repetitious reflection, a majority
of energy is lost in the useless pulses, which would result in a low
efficiency.

In this letter, we present a new setup for a single-photon qubit
against collective noise without ancillary qubits. It is made up of
three unbalanced polarization interferometers, based on a polarizing
beam splitter (PBS),  a beam splitter (BS: 50/50), and a half wave
plate (HWP). The sender encodes a qubit with time code, and the
receiver can obtain an uncorrupted state in a definite time of
arrival. The fidelity of transmission through the noisy channel is
hold by sacrificing some success percent. The success probability is
independent of the parameters of the noise and can be improved to 1
with some time delayers. Moreover, this scheme has some good
applications in quantum communication. We discuss its application in
the famous BB84 QKD for rejecting the errors caused by the
collective noise in quantum channel.

\begin{figure}[!h]
\begin{center}
\includegraphics[width=8cm,angle=0]{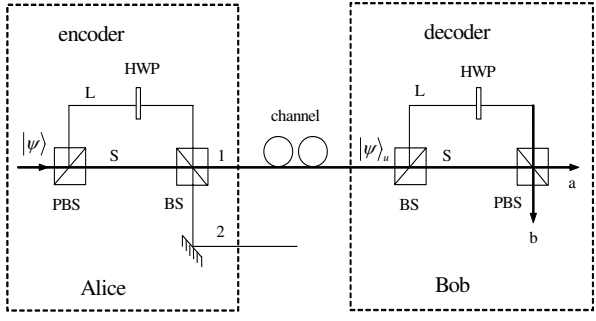}
\caption{Scheme for single-photon quantum error rejection. Alice
encodes her qubit in two time bins and sends it to Bob through a
noisy channel. Bob selects the uncorrupted state at specific time
slots. } \label{f1}
\end{center}
\end{figure}

Let us assume that the initial state to be transmitted is $\vert
\psi \rangle=\alpha \vert H \rangle + \beta \vert V \rangle$ $(\vert
\alpha^2 \vert + \vert \beta^2 \vert=1)$. Here, $\vert H \rangle$
and $\vert V \rangle$ denote the horizontal and the vertical
polarization modes of photons, respectively. Our setup is shown in
Fig. \ref{f1}. The sender Alice and the receiver Bob have the
similar set-ups,  an encoder and a decoder, including an unbalanced
polarization interferometer composed of a PBS, a BS, and a HWP. The
first PBS in Alice's side transmits the horizontal polarization mode
$\vert H \rangle$ and reflects the vertical polarization mode $\vert
V \rangle$. In this way, the state $\vert H \rangle$ propagates
through the short path (S) and the state $\vert V \rangle$ goes
through the long path (L). The time of flight difference of the
unbalanced interferometer is set to be on the order of a few
nanoseconds, which is much lesser than the time of fluctuation in
the fiber. The two proximate parts will have the same infection by
the noisy channel \cite{yamamoto}. The first HWP accomplishes the
transformation $\vert V \rangle \rightarrow\vert H \rangle$.

The state launched into the noisy channel can be described as
follows
\begin{eqnarray}
\vert \psi \rangle &=& \alpha \vert H \rangle +\beta \vert V \rangle
\nonumber\\
& \xrightarrow{{\tiny \; PBS \;}} & \alpha \vert H \rangle_S
+\beta \vert V \rangle_L \nonumber\\
& \xrightarrow{{\tiny HWP}} & \alpha \vert H \rangle_S +\beta \vert H \rangle_L \nonumber\\
& \xrightarrow{{\tiny \;\; BS \;\;}} &
\frac{1}{\sqrt{2}}\{\left(\alpha \vert H \rangle_{S} +
i\beta \vert H \rangle_{L}\right)_1 \nonumber\\
&& \,\,\,\,+ i(\alpha \vert H \rangle_{S} -i\beta \vert H
\rangle_{L})_2\}
\end{eqnarray}
The subscripts 1 and 2 represent the two output ports of Alice's BS,
shown in Fig.1, called them as  channels 1 and  2, respectively. The
coefficient $i$ comes from the phase shift aroused by the BS
reflection. Owing to the fact that the two states in two output
ports can be transformed into each other with a unitary operation,
we just consider the channel 1 in detail below, and the same way can
be used to the state coming from the channel 2 with a little
modification.

The state in the input of the channel 1 is
$\frac{1}{\sqrt{2}}(\alpha \vert H \rangle_{S} + i\beta \vert H
\rangle_{L})_1$. Notice that the polarization states of the wave
packets in the two time bins are both $\vert H \rangle$, the
influences of the noise in the quantum channel on these two pulses
are the same one. The noise of channel can be expressed with a
unitary transformation
\begin{eqnarray}
&& \vert H \rangle_j= \delta \vert H \rangle_j +\eta \vert V
\rangle_j,\label{noise}\\
 && |\delta|^2+ |\eta|^2=1,
\end{eqnarray}
$j=S,$ and $L$ denote the two time-bins. Different kinds of noises
in the quantum channel have the similar form shown in the equation
(\ref{noise}), and just are different in their parameters. The
evolution of the state from Alice to Bob through the channel 1 can
be written as
\begin{eqnarray}
&&\frac{1}{\sqrt{2}}(\alpha \vert H \rangle_{S} + i\beta \vert H
\rangle_{L}) \xrightarrow{{\tiny \;\; channel \;\;}}  \vert \psi \rangle_u \nonumber\\
&&=\frac{1}{\sqrt{2}}(\alpha \delta \vert H \rangle_{S} +\alpha \eta
\vert V \rangle_{S}+ i\beta \delta \vert H \rangle_{L}+i\beta \eta
\vert V \rangle_{L})
\end{eqnarray}
The subscript 1 was omitted for simpleness. When Bob receives the
state with the noise parameters $\delta$ and $\eta$, the action of
his unbalanced interferometer (i.e., his decoder) is given by
\begin{widetext}
\begin{eqnarray} \vert \psi \rangle_u &
\xrightarrow{{\tiny \;\; BS\;\;}} & \frac{1}{2} (\alpha \delta \vert
H \rangle_{SS} +\alpha \eta \vert V \rangle_{SS}+ i\beta \delta
\vert H \rangle_{LS}+i\beta \eta \vert V \rangle_{LS}+i\alpha \delta
\vert H \rangle_{SL} +i\alpha \eta \vert V \rangle_{SL}-\beta \delta
\vert H \rangle_{LL}-\beta \eta \vert V
\rangle_{LL}) \nonumber\\
& \xrightarrow{{\tiny HWP}} &  \frac{1}{2} (\alpha \delta \vert H
\rangle_{SS} +\alpha \eta \vert V \rangle_{SS}+ i\beta \delta \vert
H \rangle_{LS}+i\beta \eta \vert V \rangle_{LS}+i\alpha \delta \vert
V \rangle_{SL} +i\alpha \eta \vert H \rangle_{SL}-\beta \delta \vert
V \rangle_{LL}-\beta \eta \vert H\rangle_{LL}) \nonumber\\
& \xrightarrow{{\tiny \; PBS \;}} &  \frac{\delta}{2}[\alpha  \vert
H \rangle_{SS}-\beta  \vert V \rangle_{LL} + i(\underline{\beta
\vert H \rangle_{LS}+\alpha \vert V \rangle_{SL}})]_a  +
\frac{\eta}{2}[\alpha \vert V \rangle_{SS}-\beta  \vert
H\rangle_{LL} + i\underline{(\beta \vert V \rangle_{LS}+\alpha
\vert H \rangle_{SL})}]_b.\label{decode}
\end{eqnarray}
\end{widetext}
The subscripts $a$ and $b$ represent the two outputs of Bob's PBS
(see Fig. \ref{f1}). From the last line of expression
(\ref{decode}), one can see that the terms with underlines indicate
the states which will arrive at a definite time in the two outputs
$a$ and $b$. LS or SL means being delayed once, transmitting once
through S and once through L. Other states may arrive too late (LL)
or early (SS). Therefore, Bob can get the uncorrupted states in the
determinate time corresponding to SL and LS. If the photon arrives
at the output $a$, a bit flip operation is needed to get the
original state. In contrast, there is nothing needed to perform on
the photon arriving at the output $b$. The evolvement of the states
transmitting through the channel 2 is similar to that through the
channel 1. From  Eqs. (\ref{decode}), one can see that the success
probability is $\frac{1}{4}$ in  path 1, independent of the noise
parameters $\delta$ and $\eta$. Consider the contribution of the
channel 2, the total probability to obtain an uncorrupted state is
$\frac{1}{2}$.

With some time delayers Bob can obtain an uncorrupted state with the
success probability of 1 as he can also select the state at the time
bins SS and LL. In this time, he can first manipulate the relative
phases of the wave packets SS, LL and LS (SL), and then delay the
signal arriving at the time bin SS twice of the interval of the
short path (S) and the long path (L) and those at the time bins LS
(SL) once.

This quantum error rejection scheme has several important
characters. First, an arbitrary qubit error caused by the noisy
channel can be rejected with the success probability of $1$. The
significant hypothesis is that the two time bins are so near that
they suffer from the same noise infection \cite{yamamoto}. Secondly,
neither additional qubits nor entangled states are employed, so the
resource needed will not increase with the channel length. Moreover,
the setup is absolutely passive linear optics without the use of
fast polarization modulator. Thirdly, the success probability is
independent of the noise parameters. In other schemes proposed
\cite{code,correction}, the probability always relates to the noise.
They can get a certain number of success probability only by varying
the noise parameters over their ranges. Moreover, this scheme is
efficient for the transmission of a subsystem of a larger quantum
system. That is, it is also suitable for a mixed state as one needs
only to replace the state $\alpha\vert H\rangle + \beta\vert
V\rangle$ with $\alpha'\vert H\rangle_h\vert H\rangle_t +
\beta'\vert V\rangle_h\vert V\rangle_t \equiv \alpha''\vert
H\rangle_t + \beta''\vert V\rangle_t$ in the equations (1)-(5) and
obtain the same result. Here $h$ and $t$ represent the home particle
and the traveling particle, respectively.

The present scheme has some good applications in almost all one-way
quantum communication protocols \cite{rmp} for rejecting the errors
caused by the collective noise. For example, the BB84 QKD protocol
\cite{bb84} can be carried out through a noisy channel using this
time-bin encoding method. In this time, the two parties Alice and
Bob can choose the two nonorthogonal bases, X basis $\vert \pm
x\rangle=\frac{1}{\sqrt{2}}(\vert H \rangle \pm \vert V \rangle)$
and Y basis $\vert \pm y\rangle=\frac{1}{\sqrt{2}}(\vert H \rangle
\pm i\vert V \rangle)$, to prepare and measure the quantum states.
Alice products single photons randomly in one of the four states and
sends them to Bob using this error rejecting device (her encoder).
Bob first recovers the original state with his decoder and some time
delayers, and then selects one of the two bases randomly to measure
the photon. Owing to the fact that half of particles are measured
with uncorrelated bases, only 50\% of the photons are used to create
the raw key, the same as the original BB84 QKD protocol. Apparently,
the efficiency of this BB84 realization is same as that in the
original one \cite{bb84} and it rejects the errors caused by the
channel noise. Without time delayers, Bob can decode the quantum
state with success probability of $\frac{1}{2}$ and accomplish the
BB84 QKD securely, although the two parties have a low
generating-key bit rate.

In summary, we have proposed a single-photon error rejection scheme
against a collective noise with linear optics. In this scheme,
additional qubits and fast polarization modulators are not required.
Qubits are encoded in time bins and the uncorrupted state will
arrive at definite time slots. The success probability for obtaining
an uncorrupted state can be improved to 100\% with some time
delayers, which is independent of the parameters of the collective
noise. This scheme is passive and is also suitable for a mixed
state. It has some good applications in one-way quantum
communication.

This work is supported by the National Natural Science Foundation of
China under Grant Nos. 10604008 and 10435020, A Foundation for the
Author of National Excellent Doctoral Dissertation of China, and
Beijing Education Committee under Grant No. XK100270454.


\end{document}